# Visualizing the strain field in semiflexible polymer networks: strain fluctuations and nonlinear rheology of F-actin gels

J. Liu,[1] G.H. Koenderink,[1,*] K.E. Kasza,[1] F.C. MacKintosh,[2] and D.A. Weitz[1]

[1]*Dept. of Physics & DEAS, Harvard University, Cambridge, MA 02138.*
[2]*Department of Physics and Astronomy, Vrije Universiteit, Amsterdam, The Netherlands.*

* Now at: FOM Institute for Atomic and Molecular Physics (AMOLF), Amsterdam, The Netherlands.

Abstract:

We image semi-flexible polymer networks under shear at the micrometer scale. By tracking embedded probe particles, we determine the local strain field, and directly measure its uniformity, or degree of affineness, on scales of 2-100 $\mu$m. The degree of nonaffine strain depends on polymer length and crosslink density, consistent with theoretical predictions. We also find a direct correspondence between the uniformity of the microscale strain and the nonlinear elasticity of the networks in the bulk.

Actin is an abundant protein found in eukaryotic cells. *In vitro*, actin can polymerize to form semi-flexible filaments (F-actin). A myriad of actin binding proteins can both crosslink and bundle these filaments, resulting in an elastic network. Crosslinked F-actin networks are an excellent model system with which to study the physics of semi-flexible polymer networks; they can also provide insight into the mechanical properties of the cell. These networks exhibit remarkable mechanical behavior. Unlike conventional polymer gels, they are highly elastic even for small volume fractions of protein. Moreover, their elasticity can be highly nonlinear: For highly cross-linked networks, the elasticity exhibits a striking increase in stiffness with strain [1-3]; by contrast, for weakly cross-linked networks the elasticity exhibits a decrease in stiffness with strain on long time scales [4]. Understanding the origin of this behavior remains an important challenge. Several models for the microscopic origin of this nonlinear elasticity have been proposed [3, 5, 6]. Densely crosslinked networks are predicted to deform affinely, or uniformly, throughout the sample [7-9]; then network stiffening results from entropic stretching of individual semi-flexible filaments [1, 3, 5, 10]. By contrast, weakly crosslinked networks are predicted to deform nonaffinely, with a response dominated by bending of individual filaments [7-9, 11] which is not expected to lead to strain-stiffening [1]. Thus, a critical parameter to test existing models of nonlinear elasticity is the degree to which the network deforms non-affinely. Experimental measures of this require imaging the network strain field [12], and quantitatively analyzing the degree of nonaffinity.

In this Letter, we directly image the microscale strain field of sheared F-actin networks and measure the resulting nonaffinity [13]. Network deformation becomes more affine at higher crosslink density or longer filament length, which correlates directly with the simultaneous appearance of nonlinear elasticity of these networks.

To form F-actin networks and to control their structure and connectivity, we use the crosslink scruin [1, 14]. Scruin decorates the actin filaments and scruin-scruin interactions lead to the crosslinking, and, at higher concentrations, some filament bundling [15, 16]. The scruin crosslinks are both rigid and irreversible, ensuring that the network response reflects that of the actin filaments alone [1, 10, 17]. Furthermore, at low enough scruin concentration, it has been shown that the resulting networks are relatively homogeneous, making this a good model system for our study. We vary the degree of filament crosslinking by altering the molar concentration of scruin relative to actin, $R = c_S / c_A$, where $c_S$ and $c_A$ are the scruin and actin concentrations, respectively. In addition, the average contour length of F-actin, $L$, is regulated through addition of the actin severing and capping protein, gelsolin [2, 18].

We purified scruin from the acrosomal process of the *Limulus* sperm [19, 20]. Samples are prepared by gently mixing monomeric actin, scruin, gelsolin, and 10x F-buffer (20 mM Tris HCl, 20 mM $MgCl_2$, 1 M KCl, 2 mM DTT, 2 mM $CaCl_2$, 5 mM ATP, pH 7.5) at 4°C. Fluorescent carboxylate-modified latex particles of radius, $a = 0.5$ $\mu$m, are added as probes of the microscopic deformation. The solution is loaded immediately into the shear cell, surrounded by mineral oil to prevent drying, and incubated for 1 hour at 25 °C. Confocal imaging confirms that all samples are homogeneous networks without bundles (supplementary material).

We use an optical shear cell mounted on a confocal microscope (Zeiss LSM 510-Meta), enabling 3D visualization of a sample. The sample is sheared between a fixed top plate, 6 mm in diameter, and a movable microscope coverslip, as shown in the inset of Fig. 1. Set screws fix the gap to ~ 100 $\mu$m and align the plates so they are parallel to within 1 $\mu$m over the shear zone. A piezoelectric actuator moves the bottom coverslip up to 30 $\mu$m, resulting in strains of at most 30%. For larger strains, we use an alternate optical shear cell whose top plate is moved by a micrometer. We find empirically that coating both plates with bovine serum albumin prevents slipping at the surface. At each applied strain step, we first collect several images of a single plane and confirm there is no slip or drift, ensuring that we probe a purely elastic regime. We then collect a *z*-stack of 121 frames, each separated by 300 nm. There are about 200 probe particles in the sample volume and we determine the 3D positions of particle centers and monitor each of their trajectories [18, 21], with a resolution of ~ 0.02 - 0.1 $\mu$m, depending on network stiffness, and limited by thermal motion. The particle displacement in the shear direction, $\Delta y$, is linear with particle position in height, *z*, as shown in Fig. 1, where different symbols represent different shear-strain amplitudes. This confirms that the strain is, on average, affine, enabling us to determine the actual strain applied to the sample, $\gamma$, from a linear fit for each data set.

To measure the rheological properties of F-actin networks *in situ*, while the sample is prestressed by an externally applied strain, we use two-particle (2P) microrheology, which probes the elastic response of the network using correlated motion of pairs of particles; this is a robust technique particularly suitable for actin networks [18]. Because the sample is prestressed, this corresponds to a measurement of the differential or incremental modulus, $K = d\sigma/d\gamma$, where $\sigma$ is stress. For comparison, we also measure both elastic ($K'$) and viscous ($K''$) components using a stress controlled rheometer (Bohlin CVOR) with a parallel-plate geometry using a 50-$\mu$m gap. We superpose a small, oscillatory stress on a static prestress, and measure the resultant oscillatory strain to determine $K'$ and $K''$ [1, 10, 22].

The nonlinear rheology of F-actin networks is critically dependent on the degree of crosslinking of the network [1]. For a highly crosslinked network ($c_A$ = 1 mg/ml, $R$ = 0.06), the viscoelasticity probed by microrheology at all applied strains is dominated by a nearly frequency-independent $K'$ whose magnitude increases with increasing $\gamma$, as shown by the lines in Fig. 2(a). We can directly match these data to bulk measurements of $K'$ by applying increasing values of prestress, as shown by the symbols in Fig. 2(a). However, even at $\gamma = 0$, the sample measured with microrheology corresponds to a prestress of $\sigma_0 = 0.3$ Pa (squares). We speculate that this results from the induced strain on the polymerizing sample as the top plate of the shear cell is lowered into position. However, the incremental strains required to increase the applied stress on the bulk sample are identical to those applied to the microrheology sample. By contrast, for a more weakly crosslinked network ($c_A = 0.5$ mg/ml, $R = 0.002$), $K'$ probed by microrheology becomes dominant only at low frequencies and increases with frequency, but is independent of applied $\gamma$, up to 36%, as shown by the lines in Fig. 2(b). Identical results, independent of applied stress, are again obtained using bulk rheology, as shown by the symbols in Fig. 2(b).

To probe the microscopic deformation, we compare the relative positions of pairs of particles before and after shear. We show this schematically for two particles separated by distance $r$ in the inset of Fig. 3(a). The solid line represents the actual separation vector between the particles after the strain is applied while the dashed line represents this separation vector if their motion had been strictly affine. The deviation is characterized by angle $\Delta\theta$, and a corresponding distance, $r\Delta\theta$. We define the degree of nonaffinity at a length scale $r$ as $N(r) = \langle r^2 \Delta\theta^2 \rangle_r / \gamma^2$, where we average over all pairs of particles separated by $r$. This is equivalent to the method used in Ref. [7], except that our normalization is more sensitive to small changes in $\Delta\theta$.

For all samples, we find that $N(r)$ is independent of the applied strain. Given the normalization of $N(r)$, this means that the amplitude of nonaffine deformation is linear in $\gamma$. For example, for a highly crosslinked sample ($c_A = 1$ mg/ml, $R = 0.06$), $N(r)$ is identical for different strains, as shown by the overlay of the three sets of data plotted as squares in Fig. 3(b). This independence of $N(r)$ on $\gamma$ also confirms that our measurements are not influenced by imaging noise or uncertainty in position detection, which will not increase with $\gamma$. Moreover, we estimate that uncertainty due to thermal motion of the particles in the sketch in Fig. 3(a) is approximately a factor of 10 smaller than what we measure. At shorter separations $r$, however, increasing statistical uncertainty is apparent, due to the smaller number of bead pairs. We find $N(r) \sim r^{0.6}$, as indicated by the solid line in Fig. 3(b). When the crosslink density is decreased ($c_A$ = 0.75 mg/ml, R = 0.01; and = 0.5 mg/ml, R = 0.002), the magnitude of $N(r)$ increases while the exponent decreases, but the data remain independent of $\gamma$, as shown by the data sets indicated by circles and triangles in Fig. 3(b). Thus, with decreasing crosslink density, the magnitude of $N(r)$ increases while the $r$-dependence becomes weaker.

The parameter $N(r)$ measures the degree of nonaffinity at a length scale $r$. To interpret the $r$-dependence, we consider the case when $r$ is smaller than a correlation length, $l_0$. The relative deformation of two points separated by $r$ along a filament can have two limiting behaviors depending on the filament rigidity. Very rigid filaments or bundles will lead to highly correlated deformations for $r < l_0 \sim L$, since the filament will behave like a rod and rotate under shear. The deviation from affine deformation characterized by $\Delta\theta$ will then be independent of $r$, leading to $N(r) \sim r^2$ for $r < l_0$. In contrast, if the two points are not connected by a filament, or if the filament is very flexible, the points will tend to move independently, resulting in $N(r) \sim r^0$. However, this behavior will be highly sensitive to the presence of intervening filaments; as a result, we expect large sample-to-sample variations in $N(r)$ for small separations.

For larger *r*, we expect some residual correlations for densely crosslinked networks. We assume this can be roughly approximated by a power law $N(r) \sim r^\alpha$, with $0 < \alpha < 2$ when $r > l_0$. As the networks become less dense or more poorly connected, we expect the correlations to be weaker; as a result, $\alpha$ decreases toward 0. Moreover, due to a smaller energy cost for strain fluctuations in more weakly crosslinked networks, the overall magnitude of $N(r)$ will tend to increase for decreasing cross-linking over the whole range of *r*. This is summarized in Fig. 3(a). This picture is consistent with simulation results [7], which indicate a correlation length $l_0 \approx L$. While the current simulations [6-9, 23] are limited to 2D, the same qualitative behavior is expected in 3D. In our experiments, the smallest *r* we can probe is ~ 2 *μ*m. Probing smaller distances would require much smaller marker particles and therefore much better image resolution than we can currently achieve. Nevertheless, the *r* dependence of $N(r)$ we measure at larger *r*, as well as the magnitude of $N(r)$ for networks with different crosslink densities are entirely consistent with this physical picture. Specifically, we observe a decrease of *α* from 0.6 to 0.3 with decreasing crosslink density.

Decreasing *L* should produce much the same effect on the degree of nonaffinity of the networks as decreasing the crosslink density, since there will be fewer crosslinks per filament at the same molar ratio. For a fixed actin concentration ($c_A$ = 1 mg/ml) and crosslink density (R = 0.06), $N(r)$ indeed increases as *L* decreases from ~ 10 *μ*m to 1 *μ*m, through addition of increasing concentration of gelsolin, as shown in Fig. 3(c), further supporting the theoretical prediction.

To quantify the combined effect of both crosslink density and filament length on network connectivity, we use a dimensionless microstructure parameter *L*/*λ*, where $\lambda \sim l_c (l_c / l_b)^{1/3}$, with $\lambda$ the distance along a filament over which the deformation is nonaffine [7], $l_c$ the mean distance between crosslinks and $l_b$ a length of order the filament radius. Thus, the ratio $L/\lambda$ essentially measures the number of crosslinks per filament, which characterizes how strongly connected the network is. Although $l_c$ is not independently measured, previous studies of the nonlinear properties of actin-scruin networks have shown that $l_c \sim c_A^{-1/2} R^{-0.6}$ for unshortened actin filaments [1, 24]; we assume that this same dependence also applies upon addition of gelsolin. Using this, we can estimate $L/\lambda$ up to an unknown prefactor for the data shown in Figs. 3(b) and (c). Since the prefactor is not known, we scale this ratio by its value for the most weakly connected networks ($c_A$ = 0.5 mg/ml, R = 0.002). The network with the highest crosslink density and filament length has a value of $L/\lambda$ that is 13.5 times larger. Decreasing either the crosslink density or the filament length leads to a smaller $L/\lambda$, and a concomitant increase in $N(r)$.

While $N(r)$ contains information on the nonaffinity at different length scales, we also use a simplified measure giving only a single scalar quantity [25] that will facilitate comparison among many samples. We use a scalar measure of the average deviation from affinity, $S = <\vec{\Delta r} \cdot \vec{\Delta r}>/\gamma^2$, where $\vec{\Delta r}$ is the deviation vector of a single particle from the affine position after shear, as shown schematically in the inset of Fig. 4(a), and <> denotes averaging over all particles. We observe an increase in $S$ as the relative $L/\lambda$ is decreased, as shown in Fig. 4(a). This trend of $S$ is in excellent agreement with that of $N(r)$. At the same time, as the sample becomes more nonaffine with decreasing relative $L/\lambda$, the nonlinear rheology of the sample changes from stiffening (circles) to non-stiffening (triangles), consistent with the entropic stretching origin of the stiffening [1, 3, 5]. In contrast, when the network deforms more nonaffinely, there is no stiffening, since the response is dominated by bending of individual filaments [1, 7, 8], which is inherently more linear.

We have directly probed the microscopic deformation of F-actin networks, while simultaneously measuring the nonlinear rheology using microrheology. We find a correlation between bulk nonlinear rheology and microscopic structure under strain. This technique provides a promising tool to explore the origin of the striking and unusual nonlinear elastic properties of F-actin and other semiflexible polymer networks, and to examine the important role of various physiological crosslinkers such as filamin and $\alpha$-actinin that can strongly affect this nonlinear response [2, 22, 26]. While our results are qualitatively consistent with prior 2D simulations and theories [6-9, 23], 3D simulations or theory are still necessary for a deeper understanding of cytoskeletal and other biopolymer networks.

We thank T. Stossel, F. Nakamura, and G. Waller for assistance. This work was supported by the NSF (DMR-0602684), the Harvard MRSEC (DMR-0213805), and the Foundation for Fundamental Research on Matter (FOM). GHK was supported by a European Marie Curie Fellowship (FP6-2002-Mobility-6B, Contract No. 8526). The confocal microscope is maintained by the Center for Nanoscale Systems.

**Figure captions:**

FIG. 1 (Color online). Particle position *y* at height *z*, showing the displacement in the shear direction for a typical sample. Different symbols represent different strain amplitudes. The inset shows a schematic of the setup. An optical shear cell consisting of a fixed top plate and a movable bottom plate is mounted on an inverted microscope.

FIG. 2 (Color online). (a) For a strongly crosslinked network ($c_A$ = 1 mg/ml, $R$ = 0.06), the elastic differential modulus measured by 2P microrheology (lines) at different strains (solid: 0%; dash: 15%; dash-dot: 24%) is compared with that measured by bulk rheology (symbols) at different prestresses (squares: 0.3 Pa; circles: 0.5 Pa; triangles: 0.75 Pa). (b) For a weakly crosslinked network ($c_A$ = 0.5 mg/ml, $R$ = 0.002), the elastic differential modulus measured by 2P microrheology (lines) at different strains (solid: 0%; dash: 36%) is compared with that measured by bulk rheology (symbols) at prestresses 0 (squares) and 0.02 Pa (circles). All data are taken at 25 °C.

FIG. 3 (Color online). (a) Schematic summary of the physical picture described in the text. At $r < l_0$, the dense network and weak network exhibit two extreme behaviors of $N(r)$, scaling as $r^2$ and $r^0$, respectively. At $r > l_0$, the behavior is intermediate between the two extremes. The solid lines indicate the experimentally accessible region. The inset shows schematically the relative position of two particles after shear strain is applied. The solid line is the actual separation while the dashed line is the separation if the strain were strictly affine. The deviation in angle is $\Delta\theta$. (b) Degree

of nonaffinity $N(r)$ for F-actin networks with fixed filament length, 10 μm, and different crosslink density, $c_A = 1$ mg/ml, $R = 0.06$ (squares, strain 18-30%); $c_A = 0.75$ mg/ml, R = 0.01 (circles, strain 24-30%); $c_A = 0.5$ mg/ml, $R = 0.002$ (triangles, strain 23-28%). The solid line shows $\sim r^{0.6}$ and the dashed line shows $\sim r^{0.3}$. (c) $N(r)$ for F-actin networks with fixed crosslink density ($c_A = 1$ mg/ml, $R = 0.06$), and different filament length, 10 μm (squares, strain 18-30%), 1.6 μm (circles, 10-18%) and 1.1 μm (triangles, 11-20%). All data are taken at 25 °C.

FIG. 4 (Color online). Relation of (a) scalar nonaffinity parameter, $S = <\vec{\Delta r} \cdot \vec{\Delta r}> / \gamma^2$, and (b) exponent of $r$-dependence of $N(r)$ to nonlinear rheology for different relative microstructure parameter, $L/\lambda$. The stiffening and weakening samples are separated by the dotted line, and are represented by circles and triangles, respectively. For all samples, the applied strain ranges from 10-30%. In the inset of (a), $\vec{\Delta r}$ is defined as the vector connecting the actual position of a single particle after shear (circle) and its position after a strictly affine shear (dotted circle). All data are taken at 25 °C.

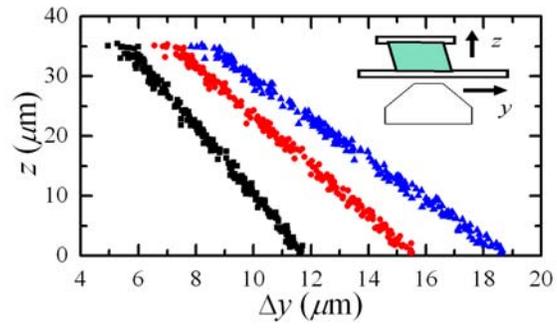

**Figure 1**

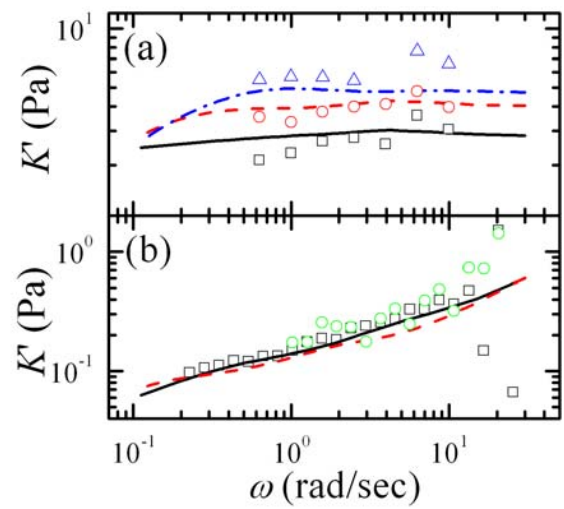

**Figure 2**

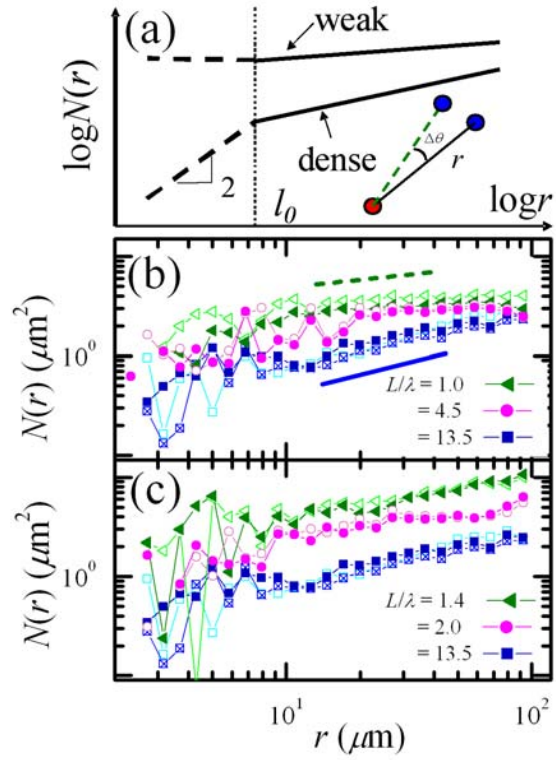

**Figure 3**

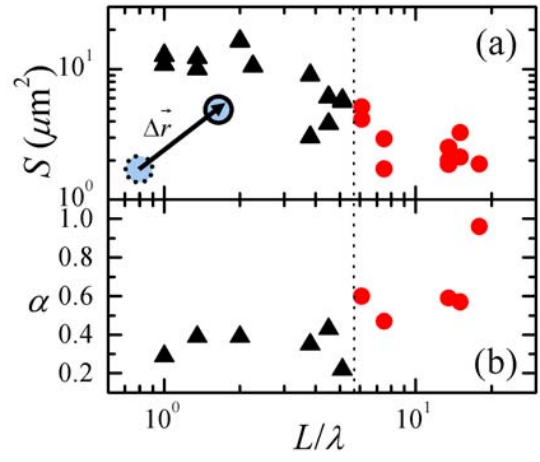

**Figure 4**

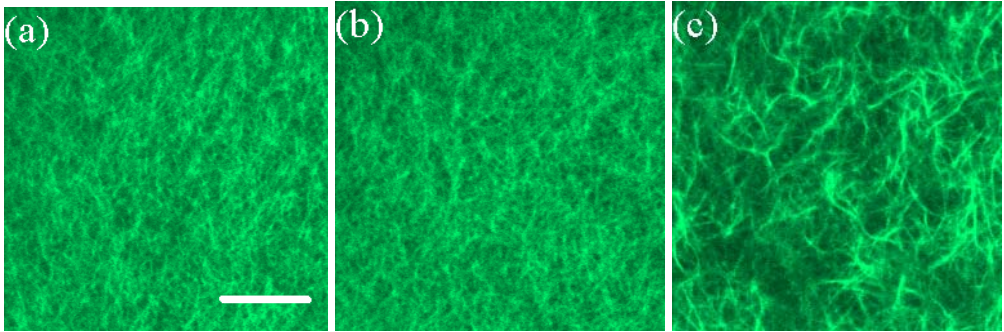

Supplementary figure:

Confocal images of actin-scruin networks with $c_A$ = 0.5 mg/ml and various $R$ values. (a) $R = 0$; (b) $R = 0.06$; (c) $R = 0.2$. Actin is labeled with Alexa phalloidin 488. The scale bar is 10 $\mu$m. Major structural change does not occur after R is increased well above 0.06. Below this limit, no significant bundling occurs and the network is homogeneous and indistinguishable from entangled actin networks. All the presented data in the paper is taken on samples in this regime.